\documentstyle[epsf,a4,11pt]{article}
\begin{document}
\hyphenation{ana-logue ana-lo-gous ana-lo-gous-ly ana-ly-sis}
\renewcommand{\baselinestretch}{1.2}

\newcommand{\deflabel}[1]{#1\hfill}%
\newenvironment{deflist}[1]%
{\begin{list}{}%
{\settowidth{\labelwidth}{#1}%
\setlength{\leftmargin}{\labelwidth}%
\addtolength{\leftmargin}{\labelsep}%
\setlength{\parsep}{0pt}%
\setlength{\itemsep}{0pt}%
\setlength{\topsep}{0pt}%
\renewcommand{\makelabel}{\deflabel}}}%
{\end{list}}%

\newcommand{\bbox}[1]{\mbox{\boldmath$#1$\unboldmath}}
\def\openone{1\hspace{-2.3pt} \mbox{l}}     

\noindent
{\LARGE \bf EPR-Bell Nonlocality, Lorentz Invariance,}

\medskip\noindent
{\LARGE \bf  and Bohmian Quantum
Theory}

\bigskip

\bigskip\noindent
{\large Karin Berndl$^{\mbox{\normalsize a}}$, Detlef
D\"urr$^{\mbox{\normalsize  a}}$,
Sheldon Goldstein$^{\mbox{\normalsize  b}}$, Nino
Zangh\`{\i}$^{\mbox{\normalsize  c}}$}

\medskip

\begin{deflist}{a }
\item[a ] Mathematisches Institut der Universit\"{a}t M\"{u}nchen,\\
Theresienstra{\ss}e 39,
80333 M\"{u}nchen, Germany
\item[b ] Department of Mathematics,
Rutgers University,\\ New Brunswick, NJ 08903, USA
\item[c ] Dipartimento di Fisica,
Universit\`a di Genova, Sezione INFN Genova,\\ Via Dodecaneso 33,
16146 Genova, Italy
\end{deflist}

\medskip\noindent
October 26, 1995

\paragraph{Abstract:}
We discuss the problem of finding a Lorentz invariant
extension of Bohmian mechanics.  Due to the nonlocality of the theory there is
(for
systems of more than one particle) no obvious way to achieve such an
extension. We present a model invariant under a certain limit of Lorentz
transformations, a limit retaining the characteristic
feature of relativity, the non-existence of absolute time resp.\
simultaneity. The analysis of this model exemplifies an important property
of any Bohmian quantum theory: the quantum equilibrium distribution $\rho =
|\psi
|^2$ cannot simultaneously be realized in all Lorentz frames of reference.

\section{Introduction}

Despite the impressive and unquestioned empirical success of quantum theory,
the
physical meaning of its basic object, the wave function, is still
controversial.
The standard---or Copenhagen---interpretation of quantum theory asserts
that the wave function embodies the most complete description possible of the
state
of a physical system, while connecting it with experience, and thereby
assigning to it physical significance, only via a set of rules for
calculating probabilities of results of ``measurements.'' It seems
essential within the standard interpretation that ``measurements'' be
distinguished from other physical processes, and that attention be paid to
the fact that the theory makes predictions {\it only}\/ about results of
``measurements'': otherwise one runs into the well-known measurement
problem or, more pictorially, the paradox of Schr\"odinger's cat.  In any
case, the fundamental role of ``measurements'' (which is sometimes shifted
to ``observers'') in the Copenhagen interpretation leads first of all to
the theory's not being well-formulated as a fundamental (as opposed to
phenomenological) theory because what constitutes a ``measurement'' is not
specified.  Secondly, with regard to cosmology, the necessity to invoke an
outside measurement apparatus or observer seems rather awkward.  (For
extraordinarily clear presentations of the problems of quantum theory as well
as of
possible solutions see \cite{Bell,Bellmeas,Albert}.)

An alternative  interpretation resp.\ theory  agreeing with quantum theory
on (most of) its predictions which is not based on the notion of
``measurement'' or ``observer'' is usually called a ``realistic''
interpretation resp.\ theory.\footnote{This is a rather
unfortunate term---can ``realism,'' i.e., the belief that there is
a material world the description of which is the task of physics,
seriously be questioned in physics? See also \cite{Maud95}.}  More precisely,
we shall
understand by a ``realistic quantum theory'' a theory,  agreeing with
quantum theory on (most of) its predictions, in which it is explicitly
specified  what the material world is thought to be made of---be
it particles or fields or what have you---and how these
entities behave. We emphasize that this by no means implies a
``naive realism''; on the contrary, these entities---what Bell called the
``beables'' of the theory---can be rather remote from our perception of the
world. Moreover, the performance of experiments may disturb
the behavior of the beables, so that the ``observed'' properties
of matter may be quite different from those left ``unobserved.''

In nonrelativistic quantum theory there are two principal routes for setting up
a realistic
quantum theory: Either the wave function is not the complete description of the
state of a
physical system, but must be supplemented by some further quantities,
commonly (and unfortunately) called ``hidden variables,'' or the unitary
evolution of the wave function must be modified. The paradigmatic example of
the first
route is Bohmian mechanics \cite{Bohm52,DGZ92a}, that of
the second route  the theory of Ghirardi, Rimini, and Weber (GRW)
\cite{GRW}. We shall call a realistic quantum theory of the first kind a
``Bohmian theory.''
Our objective is to find a Lorentz invariant Bohmian theory which extends
Bohmian mechanics,
i.e., which leads to Bohmian mechanics in the nonrelativistic limit.

\bigskip

For systems of a single particle, a Lorentz invariant Bohmian theory is
immediately specified \cite{Bohm53,BohmHiley,Holland}: the
beables are the wave function $\psi(x^\mu )$ and a particle path, which may be
specified as an integral
curve
of a 4-vector field $j^\mu$ (for example, of the current naturally
associated with the Klein-Gordon or Dirac  wave function)
\begin{equation} \label{1partbohm} \frac{dX^\mu}{ds} = j^\mu (X^\mu)
.\end{equation}
Multiplication of $j^\mu$ by a positive scalar field $a(x^\mu )$
changes only the parametrization, not the path, understood as the
equivalence class of curves $X^\mu : {\rm I\! R}\to {\rm I\! R}^4$,
$s\mapsto X^\mu(s)$ differing only in their parametrization, or as the image
$X^\mu ({\rm I\! R})$ of a curve $X^\mu$, i.e., a 1-dimensional subset of ${\rm
I\! R}^4$.
If $j^\mu$ is everywhere timelike, i.e., if $j_\mu
j^\mu >0$ with the sign convention for the  metric $g_{00}=1$,
$g_{11}= g_{22}= g_{33}=-1$,
a parametrization by proper time may be obtained by replacing Eqn.\
(\ref{1partbohm}) by
$\displaystyle \frac{dX^\mu}{d\tau } =
u^\mu (X^\mu)$ with  the 4-velocity $u=aj$, $a=(j_\mu j^\mu )^{-
1/2}$. In general
there is no distinguished parametrization, and the parametrization
chosen in writing Eqn.\ (\ref{1partbohm}) has no physical
significance as such: all equations of the form $\displaystyle \frac{dX^\mu}{ds
} =
a(X^\mu )j^\mu (X^\mu)$ with different $a$ are physically equivalent.

The Dirac current $j^\mu = \bar \psi \gamma ^\mu\psi $ is a
timelike future-oriented vector; thus the curves which are
solutions of (\ref{1partbohm}) run from $t=-\infty$ to
$t=+\infty$, never backwards in time, with velocity everywhere
bounded by $c$. In particular, every path crosses every
$t=$const.-hyperplane of every Lorentz frame of reference---or,
indeed, every spacelike hypersurface---exactly once, and thus there is
a one-to-one correspondence between paths and points---their
crossing  points---on an arbitrary spacelike hypersurface.

Because the Dirac current is divergence free, it allows moreover for a
straightforward introduction of a dynamically distinguished measure on the
set of particle paths as follows: In an arbitrary Lorentz frame, take $\rho
= j^0 = \psi ^\dagger \psi$ as the density of crossings through a
$t=t_0$-hyperplane at an arbitrary time $t= t_0$.\footnote{If $\int
j^0dx^1dx^2dx^3 <\infty$, we may normalize the measure by replacing $j$ by
$aj$ with $a^{-1}=\int j^0dx^1dx^2dx^3 $ to obtain a probability measure.}
Then
the density of crossings $\rho$ arising from (\ref{1partbohm}) satisfies
$\rho = j^0$ at all times in this frame, i.e., $j^0=\psi ^\dagger \psi$ is an
``equivariant'' density. Furthermore, ``quantum equilibrium'' $\rho = j^0$
holds then in {\it all}\/ Lorentz frames at all times. The distribution
$\rho = j^0 = \psi ^\dagger \psi$ is hence the relativistic generalization
of the ``quantum equilibrium distribution'' $\rho =|\psi |^2$ of
nonrelativistic Bohmian mechanics, which is the essential tool for the
derivation
of the
nonrelativistic quantum formalism \cite{DGZ92a}.

In fact, any divergence free current $j^\mu$, in particular also the
Klein-Gordon current which is in general not globally timelike, gives
rise to a natural measure on the set of trajectories which are integral
curves of $j^\mu$ (i.e., solutions of (\ref{1partbohm})), in a way
extending the above definition of a natural measure for the
Bohm-Dirac theory. Moreover, the fact that Klein-Gordon trajectories
possibly ``run backwards in time'' may well be viewed as naturally
describing pair creation and annihilation.
We shall discuss these topics in a subsequent work.

\bigskip

For systems of more than one particle, it is not at all obvious how to
construct a Lorentz invariant realistic quantum theory, in fact it is not even
clear
whether this
is possible at all. The problem is due to the unavoidable nonlocality of any
realistic (or, more accurately, of any precise (\cite{Bell}, pp.\ 171, 194))
version of quantum theory:
The incompleteness argument of Einstein, Podolsky, and Rosen (EPR)
\cite{EPR} together with the analysis of Bell (\cite{Bell}, Chapter
2)\footnote{For particularly clear presentations see also \cite{Bell},
Chapter 16, as well as \cite{Albert,Maudlin}.} shows that every theory
giving the quantum mechanical predictions must be nonlocal. This obviously
conflicts
with what is often considered to be the essence of Einsteinian
relativity---the locality of physical interactions. The requirement of the
Lorentz invariance of a physical theory, however, doesn't force locality. Thus
a nonlocal
Lorentz invariant theory is certainly possible. This is already rather clear
from the
meaning of the terms: While ``Lorentz invariance'' describes the behavior of a
theory
under certain transformations of reference frame, the term
``locality'' conveys that there is no action-at-a-distance. For an
exhaustive discussion, see \cite{Maudlin}. An interesting classical example
is the action-at-a-distance theory of
Schwarzschild-Tetrode-Fokker-Wheeler-Feynman (see \cite{WF} and the
references therein) replacing classical electrodynamics: In this Lorentz
invariant
theory the point charges interact directly with each other (on forward and
backward light cones)---in a manner unmediated by an electromagnetic field,
which is not a fundamental entity here.

Bohmian mechanics \cite{Bohm52,DGZ92a} is manifestly nonlocal: the
velocity of
a particle at time $t$ depends in general  upon the positions of
all the other particles at that time
\begin{equation} \label{guidlaw}
{\bf v} _k ({\bf q}_1,\dots , {\bf q}_N,t) =  \frac \hbar {m_k}
\, {\rm Im}\,  \frac {\nabla _k \psi _t({\bf q}_1,\dots , {\bf q}_N)}{
\psi _t({\bf q}_1,\dots , {\bf q}_N)} . \end{equation}
In contrast to Newtonian mechanics, where for realistic
interactions the instantaneous influence of the other particles
decreases with increasing distance, and therefore widely
separated systems are (in an certain sense) approximately
independent, for Bohmian mechanics the spatial distance between the particles
is irrelevant so long as the wave function of the entire system has a
suitably entangled form.

For a system of many Dirac particles, Bohm \cite{Bohmreview,BohmHiley}
has proposed the following guiding condition
\begin{equation} \label{bohmdirac} {\bf v} _k = \frac{ \psi^\dagger
\bbox{\alpha} _k \psi}{\psi
^\dagger \psi} , \end{equation}
which is formulated with respect to a certain reference frame, and is
in fact not  Lorentz invariant.
Analogously to the nonrelativistic theory, the quantum flux equation
which is a consequence of the many-particle Dirac equation guarantees that
$\psi^\dagger \psi$ is an equivariant ensemble density for this dynamical
system {\em in the chosen reference frame}, and therefore this theory
reproduces the quantum predictions insofar as they derive from the probability
density $\psi^\dagger \psi$. These predictions don't contain a trace of the
preferred frame: Lorentz invariance holds on the observational, but not on the
fundamental
level. (The situation is similar for Bohm's quantum field theory
\cite{Bohm52,Bohmreview,BohmHiley}.)

There have been a number of arguments to the effect that a Bohmian theory must
involve a preferred frame of reference, and thus must violate  Lorentz
invariance. The
most interesting such argument has been put forward by Hardy \cite{Hardy},
who by discussing an intriguing experiment---one that we shall discuss in
this paper as well, and that has been shown to contain even more surprises
(\cite{Hardymax,Shellymax}, and in particular a nonlocality argument in a sense
involving but one photon \cite{onephot})---claims to have shown that every
realistic quantum theory must possess a preferred frame of reference, and thus
that there can be no Lorentz invariant realistic quantum theory.

However, because it rests on an unsuitable ``reality criterion''
\cite{Comment,Hreply}, Hardy's argument is wrong.  There are even
counterexamples to Hardy's argument: the multitime translation invariant
formulation of
the GRW theory by Bell (\cite{Bell}, Chapter 22) as well as the multitime
translation
invariant Bohmian theory we
present in this paper are realistic models for the discussed experiment
without a preferred frame.  Furthermore, there is an outline for a relativistic
Bohmian quantum field theory, in which a foliation of space-time into
spacelike hypersurfaces is an additional beable \cite{DGZ90}. Finally one
can find a number of models of relativistic $N$-particle theories with an
action-at-a-distance defined in a Lorentz invariant manner, models that
therefore have
the potential to properly and relativistically describe quantum
nonlocality as exhibited in Hardy's experiment. We allude to one such
possibility in Section \ref{4}, but shall discuss these models in a
subsequent work.  No nontrivial Lorentz invariant realistic quantum theory is
as yet known,
but there is no
compelling argument that this should be impossible.\footnote{And the history
of the
issue of hidden variables, i.e., of the completeness of the description
provided by the wave function, should strongly warn us against too readily
accepting impossibility claims.}  On the contrary, the above mentioned
models are steps towards a Lorentz invariant realistic quantum theory.
One should, however, be aware that the determination of the empirical
predictions of these models may present a difficult problem; in fact, for many
models there is in
general no reason that quantum equilibrium should hold with respect to any
reasonable family of hypersurfaces; thus the statistical analysis will be
different from that in nonrelativistic Bohmian mechanics and moreover,
presumably, the
predictions of such a theory won't agree with (all of) those of quantum theory.

Similarly Albert (\cite{Albert}, p.\ 159ff),  Bohm and Hiley
(\cite{BohmHiley}, Section 12.6), Ghirardi et al.\
\cite{Ghirpar}, and Hardy and Squires \cite{HarSqu} also argue that a
Bohmian theory must violate Lorentz invariance because a preferred frame
is needed.  The above mentioned models without a
preferred frame (but with some ``simultaneity'' fixed in a Lorentz invariant
way---note that this entails that there always are Lorentz frames
in which future events influence the past, in contrast to assumptions in
\cite{Albert,BohmHiley,Ghirpar,HarSqu}) show that less is
established than claimed.

\bigskip

This paper is organized as follows: We show in Section \ref{2}  that the
joint distribution of the particle positions cannot in general
agree with the quantum mechanical distribution in all Lorentz frames. This is
in contrast to the situation for 1 particle---or, indeed, $N$
independent particles---as explained above. We also discuss why
nevertheless the quantum mechanical predictions for performed measurements can
be obtained. In Section \ref{3} we present a concrete step towards a Lorentz
invariant Bohmian theory: a Bohmian theory
invariant under  certain limits of Lorentz transformations, limits defining a
symmetry that expresses the essence of relativistic space-time---the
non-existence
of absolute time resp.\ simultaneity. These transformations, which we shall
call ``multitime translations,'' have been discussed by Bell in
connection with the GRW theory (\cite{Bell}, Chapter 22, and
\cite{Belltalk}; Bell calls them ``relative time translations''). In Section
\ref{3.1} we describe a multitime translation invariant formulation of
Schr\"odinger's equation for systems composed of noninteracting parts. In
Section \ref{3.2} we present the corresponding multitime translation invariant
Bohmian theory and discuss its statistical properties. In Section \ref{3.3} we
apply the general analysis to Hardy's experiment, focusing on how this
experiment illustrates the general discussion in Section \ref{2}.

We remark that
there is no difficulty formulating a Lorentz invariant multitime version
of the Dirac equation for a system of noninteracting Dirac particles
\cite{wentzel}. However, the corresponding Lorentz invariant Bohmian theory
lacks
statistical transparency. Indeed, at first sight, Lorentz invariance and
statistical
transparency appear to be mutually exclusive.
See Section \ref{4} for a bit more detail on this, as well
as some further reflections on Lorentz invariance.

For systems that consist of noninteracting subsystems, Bell has shown that
the GRW theory can be reformulated in such a way that it becomes invariant
under multitime translation (\cite{Bell}, Chapter 22). He regarded this as an
important step towards
a genuinely
Lorentz invariant precise formulation of quantum theory, declaring that ``And I
am
particularly struck by the fact that the model is as Lorentz invariant as
it could be in the nonrelativistic version. It takes away the ground of my
fear that any exact formulation of quantum mechanics must conflict with
fundamental Lorentz invariance.''  (\cite{Bell}, p.\ 209). The multitime
translation
invariant Bohmian theory we
discuss in this paper may, perhaps, be regarded as showing that this assertion
applies also to Bohmian mechanics.

\bigskip

For simplicity, we shall put all masses $m_k=1$ and $\hbar =c=1$.

\section{Quantum equilibrium cannot hold in all Lorentz frames}\label{2}

We consider an arbitrary theory for $N(\geq 2)$ particles,
i.e., a (possibly statistical) specification of all possible $N$-tuples of
space-time paths for the $N$ particles (for example as given by solutions
of a system of differential equations). We shall call each such possible
``history'' an {\it N-path}.  We assume that each spacelike hypersurface
is crossed exactly once by each trajectory, and consider an arbitrary
probability measure $P$ on the $N$-paths. This determines the distribution of
crossings $\rho ^\Sigma :\Sigma ^N \to {\rm I\! R}$ for any spacelike
hypersurface $\Sigma$.

We now want the probabilistic predictions of the theory to agree as far as
possible with those of quantum theory. Complete agreement would be
straightforward if for any quantum state $\psi$ there were a $P$ such that
for all spacelike hypersurfaces $\Sigma$ the distribution of crossings $\rho
^\Sigma$
agrees with the quantum mechanical joint distribution of the (measured)
positions on
$\Sigma$. For $\Sigma$ a spacelike hyperplane, i.e., a simultaneity plane or
constant-time slice of a Lorentz frame $\Lambda$, this is given by
${|\psi^{\Sigma}|}^2$ where $\psi^{\Sigma}=\psi^{\Lambda}$, the wave
function in
frame $\Lambda$.  However, this is not in general possible:

\medskip

\begin{deflist}{$(*)$}
\item[$(*)$]{\it There does not in general exist a probability measure P
on N-paths for which the  distribution of crossings  $\rho^\Sigma$ agrees
with the corresponding quantum mechanical  distribution  on all spacelike
hyperplanes
$\Sigma$.}
\end{deflist}

\medskip

The field theoretical analogue of this assertion has been
conjectured by D\"urr, Goldstein, and Zangh\`{\i} in 1990
\cite{DGZ90}. Samols discusses the equivalent result for his
stochastic realistic model of a light cone lattice quantum field
theory \cite{Samols}.

The caveat ``in general'' refers to the fact that there are exceptional
physical situations for which such a $P$ does exist. Consider, for example, 2
independent Dirac particles, i.e., with a wave function that is a product of
1-particle wave functions $\psi = \psi _a \psi_b$ and independent evolutions
given by
(\ref{1partbohm}): $\frac
{dX_k}{ds}=j_k(X_k)$, $j_k^\mu =\bar \psi _k \gamma ^\mu \psi _k$,
$k=a,b$. Then, as explained above, if $\rho^{\Sigma_0} = j^0_a j^0_b$ with
respect to one spacelike hyperplane $\Sigma_0$, then $\rho ^\Sigma = j^0_a
j^0_b$ for all spacelike hyperplanes $\Sigma$. We believe, however, that
such exceptional physical situations are rare.

The assertion $(*)$ is more or less an immediate consequence of any of the
no-hidden-variables-nonlocality theorems---Bell's \cite{Bell}, that of Clauser,
Horne, Shimony, Holt \cite{CHSH}, that of Greenberger, Horne, Zeilinger
\cite{GHZ} (see also \cite{Mermin}), or
what have you---for the spin components of a
multiparticle system: By means of a suitable placement of appropriate
Stern-Gerlach  magnets the inconsistent joint spin correlations can be
transformed to (the same) inconsistent joint spatial correlations for
particles at different times. Since the existence of a probability measure $P$
on $N$-paths implies the existence and hence the consistency of all
crossing distributions, the assertion  follows.

Since this this is an important result, we shall provide an elaboration
using one of the sharpest nonlocality theorems, that of Hardy
\cite{Hardy}. It should be clear from our treatment of this example how to
arrange the magnets to deal with any other version.

\begin{figure}[t]
\begin{center}
\leavevmode
\epsfxsize=10.3cm
\epsffile{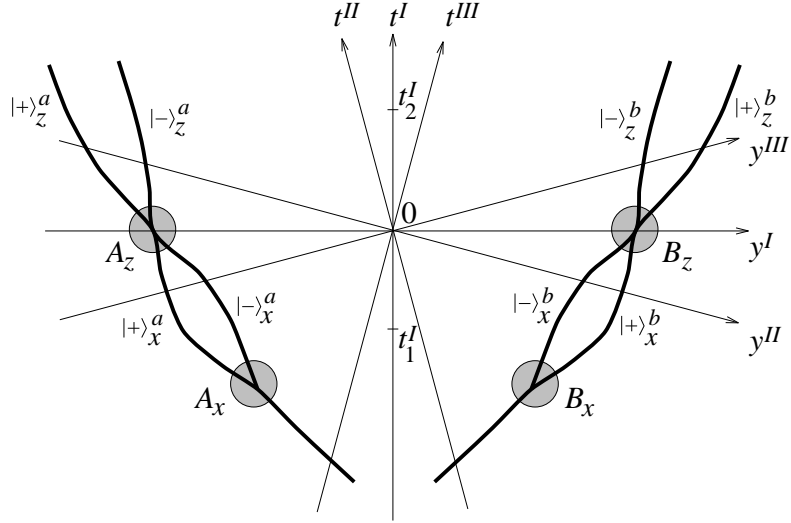}
\end{center}
\caption{Space-time diagram of the evolution of the  wave function in
Hardy's experiment. In the shaded regions there are Stern-Gerlach
magnets $A_x$, $A_z$, $B_x$, and $B_z$, which split the respective
parts of the wave function into the respective eigenfunctions $({|+\rangle ,|-
\rangle })_{x,z}^{a,b}$. Three different frames of
reference are also drawn.} \label{rhopsibild}
\end{figure}

Consider the experiment described in  Figure \ref{rhopsibild}, which is
similar to the EPR-Bohm experiment and which is a slight
modification of the experiment discussed by Hardy \cite{Hardy}, which we
shall call ``Hardy's experiment.'' A pair of particles is
prepared in Hardy's state $\psi =\psi _{\rm Hardy}$, which has, say  in frame
{\it I},  the form (we write only
the ``spin'' part)
\begin{eqnarray}  \psi_{\rm Hardy} & =  & \frac 1{\sqrt 3} \Bigl(
|+\rangle _z^a|-\rangle _z^b-\sqrt 2 |-\rangle _x^a|+\rangle _z^b
\Bigr) \label{hardy1}\\
& =  & \frac 1{\sqrt3} \Bigl( |-\rangle _z^a|+\rangle _z^b - \sqrt
2 |+\rangle _z^a|-\rangle _x^b \Bigr)  \label{hardy3}\\
&  = & \frac 1{\sqrt 3}\Bigl( |+\rangle _z^a|-\rangle _z^b-
|+\rangle _z^a|+\rangle _z^b+|-\rangle _z^a|+\rangle _z^b \Bigr)
\label{hardy2}\\
& = &   \frac 1{\sqrt{12}} \Bigl( |+\rangle _x^a|+\rangle
_x^b - |+\rangle _x^a|-\rangle _x^b - |-\rangle _x^a|+\rangle _x^b
- 3 |-\rangle _x^a|-\rangle _x^b \Bigr), \label{hardy4}
\end{eqnarray}
where $|+\rangle _x,|-\rangle _x$ denote the eigenfunctions of $\sigma
_x$ with eigenvalue $+1$ resp.\ $-1$, and $|+\rangle _z,|-\rangle
_z$ denote the eigenfunctions of  $\sigma _z$ with eigenvalue $+1$
resp.\ $-1$. We have used that  $|+\rangle _x=(|+\rangle _z+|-
\rangle _z)/\sqrt 2$ and $|-\rangle _x= (|+\rangle _z-|-\rangle
_z)/\sqrt 2$.
Denoting by $(a,b)_{(x,z)}$ the components of spin in
direction $x$ resp.\ $z$ of particle $a$ resp.\ $b$, the following quantum
mechanical
predictions can be read off from the form of the wave function:
\begin{eqnarray} & a_x=+1 \ \Rightarrow \ b_z=-1 & (\mbox{from }
(\ref{hardy1}) ) \label{pred1}\\
& b_x=+1 \ \Rightarrow \ a_z=-1  & (\mbox{from }  (\ref{hardy3}) )
\label{pred3}\\
 & \mbox{not }(a_z=-1 \ \mbox{and} \  b_z=-1)  & (\mbox{from }
(\ref{hardy2}) ) \label{pred2}\\
& \displaystyle \mbox{Prob}(a_x=+1 \ \mbox{and} \  b_x=+1)= \frac 1{12} \quad &
(\mbox{from }(\ref{hardy4}) ) \label{pred4}
\end{eqnarray}
These predictions are clearly inconsistent for random variables since the
last one together with the first two then imply that $\{a_z=-1$
and $b_z=-1\}$ has probability at least 1/12.

Now suppose that the setup is such that after the two particles are widely
separated from each other, each of them runs through a Stern-Gerlach magnet
$A_x$ resp.\ $B_x$, which splits the respective parts of the wave function into
the
eigenfunctions $|+\rangle _x^a$ and $|-\rangle _x^a$ resp.\ $|+\rangle
_x^b$ and $|-\rangle _x^b$. These parts are  later recombined by reverse
magnets  after which they are lead through a second Stern-Gerlach magnet
$A_z$ resp.\ $B_z$,
which splits the wave function into the eigenfunctions $|+\rangle _z^a$ and $|-
\rangle
_z^a$ resp.\ $|+\rangle _z^b$ and $|-\rangle _z^b$. Thus the spin
components are (more or less) perfectly correlated with the path variables
as indicated in Figure \ref{rhopsibild}, which therefore inherit the
inconsistency of the spin components. The assertion follows.

We remark that the measurements to which the quantum mechanical predictions
refer might well be performed in this way, but with the insertion of
photographic plates behind the appropriate Stern-Gerlach magnets.

We perhaps should be even more explicit, particularly since we will need
later to refer to some of the notation to be developed here. Suppose that
there is a theory for 2 particles for which the distributions of
crossings $\rho^\Sigma$ agrees with the quantum mechanical distribution for
position
(measurements),
given by $|\psi ^\Sigma |^2$, for all spacelike hyperplanes $\Sigma$. The
hyperplanes we shall consider are simultaneity planes in the
Lorentz frames {\it I} at $t^I=t^I_1$ and $t^I_2$, {\it II} at $t^{\it
II}=0$, and {\it III} at $t^{\it III}=0$, as shown in Figure
\ref{rhopsibild}. We shall denote these by $\Sigma ^{\it I}(t^I_1)$, $\Sigma
^{\it
II}(0)$, etc.. Furthermore we shall abbreviate $\psi ^{\Sigma ^{\it I}(t^I_1)}$
by $\psi ^I_1$, $\rho ^{\Sigma ^{\it I}(t^I_1)}$ by $\rho ^I_1$, $\psi ^{\Sigma
^{\it II}(0)}$ by $\psi ^{\it II}_0$, $\rho ^{\Sigma ^{\it II}(0)}$ by $\rho
^{\it II}_0$, etc..

Consider the configurational part of the wave function in these Lorentz frames.
We
shall now regard $|\pm \rangle ^{a,b} _{x,z}$ as representing the
appropriate configurational part of the wave function as indicated in Figure
\ref{rhopsibild}, with ${\rm supp}\,  |\pm \rangle ^{a,b} _{x,z}$ denoting its
spatial
support.\footnote{It should perhaps be noted that a Dirac spinor which in frame
{\it
I}\/ is a spin $x$/$z$ eigenfunction will not be a spin $x$/$z$
eigenfunction in the frames {\it II}\/ or {\it III}\/ which are boosted in
the $y$-direction. Our notation here should not be construed as implying
otherwise.} Then
\begin{eqnarray}
\psi^I_1 & = &   \frac 1{\sqrt{12}} \Bigl( |+\rangle _x^a|+\rangle
_x^b - |+\rangle _x^a|-\rangle _x^b - |-\rangle _x^a|+\rangle _x^b
 - 3 |-\rangle _x^a|-\rangle _x^b \Bigr)
\label{psi11} \\
\psi ^{\it II}_0 & =  &   \frac 1{\sqrt6} \Bigl( |-\rangle
_z^a|+\rangle _x^b + |-\rangle _z^a|-\rangle _x^b - 2 |+\rangle
_z^a|-\rangle _x^b \Bigr) \label{psi2} \\
\psi ^{\it III}_0 & =  & \frac 1{\sqrt 6} \Bigl( |+\rangle _x^a|-
\rangle _z^b + |-\rangle _x^a|-\rangle _z^b -  2 |-\rangle
_x^a|+\rangle _z^b \Bigr)  \label{psi3} \\
\psi ^I_2 & = &  \frac 1{\sqrt 3}\Bigl( |+\rangle _z^a|-\rangle
_z^b-|+\rangle _z^a|+\rangle _z^b+|-\rangle _z^a|+\rangle _z^b
\Bigr)  \label{psi12}
\end{eqnarray}

{From} the assumption that $\rho ^\Sigma = |\psi ^\Sigma |^2$ in all
frames, we obtain, from (\ref{psi11}) or (\ref{pred4}), that for the
simultaneity
surface  $\Sigma = \Sigma ^{\it I}(t^I_1)$
\begin{eqnarray} & & \int\limits_{{\rm supp}\,  |+\rangle _x^a \times {\rm
supp}\,  |+\rangle
_x^b} \rho^I_1(q_a,q_b)\, dq_a\, dq_b  \nonumber\\ & &  = \int\limits_{{\rm
supp}\,
|+\rangle _x^a \times {\rm supp}\,  |+\rangle _x^b}
|\psi ^I_1(q_a,q_b)|^2 \, dq_a\, dq_b  =  \frac 1{12} . \label{wid1}
\end{eqnarray}
For the simultaneity surfaces  $\Sigma ^{\it II}(0)$ and  $\Sigma ^{\it
III}(0)$ we have from (\ref{psi2}) resp.\ (\ref{psi3}) (or
(\ref{pred3}) resp.\ (\ref{pred1})) that
\begin{eqnarray} \rho^{\it II}_0 (q_a, q_b)=0  & \mbox{ for } & (q_a,
q_b) \in {\rm supp}\,  |+\rangle _z^a \times {\rm supp}\,  |+\rangle _x^b
,\label{wid2} \\
\rho^{\it III}_0 (q_a, q_b)=0  & \mbox{ for } & (q_a, q_b) \in {\rm supp}\,
|+\rangle _x^a \times  {\rm supp}\,  |+\rangle _z^b , \label{wid3}
\end{eqnarray}
and  for $\Sigma = \Sigma ^{\it I}(t^I_2)$ from (\ref{psi12}) or
(\ref{pred2}) that
\begin{equation} \label{wid4} \rho^{\it I}_2 (q_a, q_b)=0 \ \ \mbox{ for } \ \
(q_a,
q_b) \in {\rm supp}\,  |-\rangle _z^a\times {\rm supp}\,  |-\rangle _z^b
.\end{equation}
Consider now that part of the ensemble of two-paths containing paths that cross
${\rm supp}\,  |+\rangle _x^a \times {\rm supp}\,  |+\rangle _x^b $. {From}
(\ref{wid1}), this has probability $1/12$. {From} (\ref{wid2}), particle
$a$ will be in ${\rm supp}\, |-\rangle _z^a $ at $t^I_2$; from  (\ref{wid3}),
particle $b$ will be in ${\rm supp}\, |-\rangle _z^b $ at  $t^I_2$; thus
\[ \int\limits_{{\rm supp}\,  |-\rangle _z^a\times {\rm supp}\,  |-\rangle
_z^b}
\rho^I_2 (q_a,q_b)\, dq_a\, dq_b  \geq  \frac 1{12}, \]
in contradiction with (\ref{wid4}). (This argument assumes that, say, the
crossing track of particle $a$ for $\Sigma ^{\it I}(t^I_2)$ agrees with that
for $\Sigma ^{\it II}(0)$---i.e., that there is no sudden change of track. By
a suitable choice of geometry the violation of this assumption can be made
as implausible as we like.)
\hspace*{\fill} $\Box$

\bigskip

We can more briefly, though somewhat imprecisely, rephrase $(*)$ by saying
that ``Quantum equilibrium ($\rho=|\psi |^2$) cannot hold in all Lorentz
frames.''  Although the
notions of the wave function $\psi$ in position representation as well as that
of
a
position measurement are
problematical in relativistic quantum theory, the impact of  this statement is
not thereby
diminished.  In fact, the statement ``$\rho=|\psi |^2$ cannot hold
in all Lorentz frames'' should be understood as follows: The joint
distributions
given by  quantum theory for position measurements (from whatever formalism
they arise)
cannot in general agree with the distribution of the actual particle positions
in
all Lorentz frames.  This is the case, as pointed out above, already if
only the (experimentally well-established) predictions of the distribution of
spin measurements---spin is measured, as is any observable, cf.\
\cite{DDGZ}, ultimately by measuring some position in a suitable experiment
(here with Stern-Gerlach magnets)---in the singlet state (which is the
relevant state for the earlier versions of the nonlocality theorems) are
considered.

An immediate question is whether this leads, for a theory with
trajectories, to experimentally detectable violations of quantum mechanical
predictions. That it ain't necessarily so will be illustrated by a concrete
model in a later section. But it is already clear from (nonrelativistic)
Bohmian mechanics that
the validity of $\rho = |\psi |^2$ in just {\it one}\/ frame is sufficient
to derive the quantum mechanical predictions for observations at different
times:
Assume that the frame corresponding to (Newtonian) absolute time---the
frame in which quantum equilibrium $\rho=|\psi |^2$ holds for Bohmian
mechanics---corresponds
to system
$I$ in Hardy's experiment in Figure \ref{rhopsibild}. To derive from Bohmian
mechanics
the correct prediction for the joint distribution of a measurement of $a_x$
and a later measurement of $b_z$, one has to take into account that the
actual performance of measuring $a_x$, which requires an intervention such
as the suitable insertion of a photographic plate, influences the future
evolution of the whole system, and in particular,  nonlocally and
instantaneously, the future path of particle $b$. This can be conveniently
described in terms of the effective ``collapse of the wave function.''  The
``unmeasured'' distributions do not in general give the correct predictions for
the outcomes of experiments! For a rather detailed discussion of related
matters, see \cite{DGZ92a}, Sections 8--10.

Moreover, it is rather clear that any two theories agreeing at all times on
the  spatial distribution of particles for some frame must be
empirically equivalent, though we shall not try here to give a precise
formulation of this assertion. We note, however, that for a theory
involving a foliation of space-time into hypersurfaces, such as the proposal of
D\"urr, Goldstein, and Zanghi \cite{DGZ90}, as well as that of Samols
\cite{Samols}, it is natural to demand that ``quantum equilibrium'' hold on
these
hypersurfaces. For the proposal
in section \ref{3.2} of this paper, a theory involving particle interactions
that are
instantaneous with respect to a specified synchronization,
one is lead to demand ``quantum equilibrium'' with
respect to this synchronization.  That this indeed suffices to recover the
quantum mechanical predictions for the outcomes of all joint measurements is
implied
by the fact that the joint results for any family of measurements can
always be transferred to a common place and time---and must be if these
results are to be subject to the analysis of a single individual (cf.\
\cite{DGZ90,Samols}, and \cite{DGZ92a}, point 19 on p.\ 900). This suggests
that even a suitable kind of ``local  quantum equilibrium'' should be
sufficient to obtain
the standard quantum mechanical predictions.

\section{The multitime formalism}\label{3}

\subsection{Multitime translation invariance}\label{3.1}

\begin{figure}[t]
\begin{center}
\leavevmode
\epsfxsize=7.2cm
\epsffile{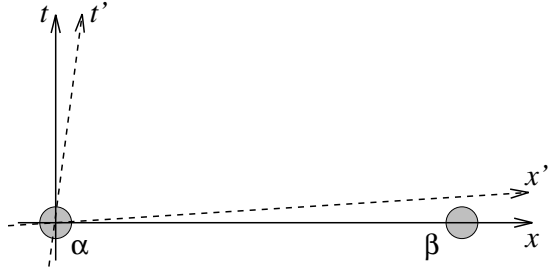}
\end{center}
\caption{$\alpha$ and $\beta$ are two widely separated events. In the
primed frame of reference, corresponding to a slowly moving
observer, these two events are not simultaneous.} \label{trafobild}
\end{figure}

Consider a system composed of $n$---we put $n=2$ for simplicity---widely
separated subsystems. Even observers who are slowly (``nonrelativistically'')
moving relative to each other need not agree on the simultaneity of events
in the separated subsystems: let $(t_\alpha , x_\alpha )$, $(t_\beta , x_\beta
)$
be the
coordinates of the events $\alpha$ resp.\ $\beta$ for observer 1.  We may put
$t_\alpha=0$, $x_\alpha=0$.  The two events are simultaneous,
$t_\alpha=t_\beta$, and
widely separated from each other, $x_\beta \gg 1$. A second observer, slowly
moving in the $x$-direction relative to the first observer, will describe
the same events by the following primed coordinates, cf.\ Figure
\ref{trafobild}:
\begin{eqnarray*} t_\alpha'=t_\alpha=0, &  & x_\alpha'=x_\alpha=0,\\
 t_\beta'=\gamma (t_\beta-vx_\beta) \approx -\vartheta, &  &
x_\beta' =\gamma (x_\beta-vt_\beta) \approx x_\beta,
\end{eqnarray*}
where $v\approx 0$, so that $\gamma = 1/ \sqrt{1-v^2} \approx 1$. It is
further assumed that $x_\beta$ is sufficiently large that $vx_\beta =\vartheta$
is
of
order unity.  For observer 2, the events $\alpha$ and $\beta$ are not
simultaneous, $t_\alpha' \neq t_\beta'$, not even approximately. More
precisely,
in the limit in which $x_\beta \to \infty$ and $v\to 0$ in such a manner
that $vx_\beta =\vartheta \neq 0$, the Lorentz transformation becomes simply a
translation of
relative time.  Consequently, for the case of a system composed of widely
separated subsystems we might demand of a nonrelativistic theory invariance
with respect to independent shifts of the zeros of the subsystems' time
scales (on subsystem clocks).  The relevance of this nonrelativistic
residue, or analogue, of Lorentz invariance, especially for the discussion of
the
possibility of a Lorentz invariant realistic quantum theory, has been pointed
out by
Bell (\cite{Bell}, Chapter 22, and \cite{Belltalk}).

To specify the space-time transformation corresponding to this
change in frame of reference, we have to introduce two separate
coordinate systems for the two widely separated subsystems $a$ and
$b$. On configuration-space-time, the multitime translation
is given by
\begin{eqnarray} L_\tau :\  {\rm I\! R}\times {\rm I\! R}^{3N_a} \times {\rm
I\! R}\times
{\rm I\! R}^{3N_b}
& \longrightarrow & {\rm I\! R}\times {\rm I\! R}^{3N_a}
\times {\rm I\! R}\times {\rm I\! R}^{3N_b}, \ \tau = (\tau_a,\tau_b) \in {\rm
I\!
R}^2 \nonumber\\  z:=(z_a,z_b):=(t_a,q_a,t_b,q_b) & \longmapsto &
(t_a-\tau_a,q_a,t_b-\tau_b
,q_b)=z'=L_\tau  z
\label{Ltaudef}
\end{eqnarray}
where $N_a$ and $N_b$ are the particle numbers of the respective
subsystems.

At first thought, one might not expect a  quantum theory to be
invariant under $L_\tau$, because absolute time seems necessary to mediate
the action-at-a-distance of Schr\"odinger's equation, not to mention the
more explicit nonlocality of Bohmian mechanics.  Indeed, for the usual
Schr\"{o}dinger equation as well as for the GRW model and Bohmian mechanics it
would
appear that the
multitime translation cannot be discussed at all because time appears in the
wave function only as common (absolute) time.

But if the subsystems $a$ and $b$ are independent, i.e., if there
is no interaction potential between the subsystems
\begin{eqnarray*} & V(q_a,q_b) = V_a(q_a) + V_b(q_b) & \\
&  H=H_a+H_b,\quad  H_k= -\frac 12 \Delta _k + V_k ,\quad k=a,b &
\end{eqnarray*}
so that the Hamiltonians $H_a$ and $H_b$ commute,
the Schr\"odinger evolution may be reformulated so that it becomes  multitime
translation invariant: {From} the ordinary one-time wave function $\psi _t=e^{-
iHt}\psi_0=U_t\psi_0$ we
define a two-time wave function
$\psi (t_a, t_b)\in L^2({\rm I\! R} ^{3N_a})\otimes L^2({\rm I\! R} ^{3N_b})
\cong  L^2({\rm I\! R} ^{3(N_a+N_b)})$
\[ \psi (t_a,t_b) = e^{-iH_at_a} e^{-iH_bt_b} \psi _0=U_{t_a}^a U_{t_b}^b
\psi_0 \]
satisfying  two separate Schr\"{o}dinger equations
\begin{equation} \label{sgl2} i\, \frac{\partial \psi}{\partial {t_a}} =H_a
\psi,\quad  i\,
\frac{\partial \psi }{\partial {t_b}} = H_b \psi . \end{equation}
This system of partial differential equations, with $\psi$ transforming in
the obvious way,
\[ \psi (z)  =  \psi \circ L_\tau ^{-1} (L_\tau z) =: \psi '(z')
 \] is invariant under $L_\tau$. In particular, the unitary representation of
the
group of multitime translations is given by $U_\tau = U^a_{\tau_a}U^b
_{\tau_b}$:
\begin{equation} \label{Utaudef} \psi ' = e^{-iH_a\tau_a }e^{-iH_b\tau_b }\psi
= U^a
_{\tau_a}U^b _{\tau_b} \psi = U_\tau \psi \end{equation}

Note that in any frame of reference given by a particular synchronization of
the
subsystem times, i.e., where $t_a=s$ is simultaneous with $t_b=s+h$,  the wave
function
in ``frame $h$,'' which is given by $\psi^{h}_t=\psi(t,t+h)$ and recognized as
$\psi^h_t= \psi '(t,t)$ for a multitime shift by $\tau = (0,h)$ from
the unprimed frame, satisfies the one-time  Schr\"{o}dinger equation.

It is also easy to see that the transition to a two-time wave function
transforms the usual quantum measurement formalism into a multitime translation
invariant
form. We shall use here the Heisenberg picture for convenience as well as for
analogy with relativistic quantum theory. Let $\psi =\psi_0=\psi (0,0)$ be the
Heisenberg
state of the system, and
consider a sequence of observables $(M^a_j)_{1\leq j\leq k}$ and
$(M^b_j)_{1\leq j\leq \ell }$, which are measured at times
$t^a_1<\dots < t^a_{k}$ resp.\ $t^b_1<\dots <t^b_{\ell}$.  Here
$M^a_j$ acts only on system $a$, i.e., $M^a_j= O^a_j\otimes \openone$ with
observables $O^a_j$ on $L^2({\rm I\! R} ^{3N_a})$, and $M^b_j$ acts only
on system
$b$, i.e., $M^b_j= \openone \otimes O^b_j$ with observables $O^b_j$ on
$L^2({\rm I\! R} ^{3N_b})$.  Thus the observables and the unitary evolution of
system $a$, $M^a_j$ and $U^a_{t_a}=e^{-iH_at_a}$, commute with the
observables and the unitary evolution of system $b$, $M^b_j$ and $
U^b_{t_b}=e^{-iH_bt_b}$: for all $j,j',t_a,t_b,$
\begin{equation} [M^a_j,M^b_{j'}]=0, \quad [M^a_j, U^b_{t_b}]=0,
\quad [M^b_j,
U^a_{t_a}]=0, \quad [U^a_{t_a}, U^b_{t_b}]=0.  \label{commute}
\end{equation}
We shall assume for simplicity that all the observables $M^a_j$ and $M^b_j$
have discrete spectrum and denote by $\pi ^a_{j,\alpha }$ resp.\ $\pi
^b_{j,\beta
}$ the projection operator onto the eigenspace of $M^a_j$ resp.\ $M^b_j$
corresponding to the eigenvalue $\alpha$ resp.\ $\beta$.  We introduce the
Heisenberg operators \[ \pi^a_{j,\alpha}(t_a)  :=  U^a_{-
t_a}\, \pi
^a_{j,\alpha}\, U^a_{t_a} \quad \mbox{and} \quad \pi^b_{j,\beta}(t_b)  :=
U^b_{-t_b}\, \pi ^b_{j,\beta}\, U^b_{t_b}, \]
which, by (\ref{commute}), agree with the usual ones involving the full
evolution $U_t=U^a_tU^b_t$.
The joint probability for obtaining the measurement results $M^a_1=\alpha_1$,
\dots
,  $M^a_{k}=\alpha_{k}$,
$M^b_1=\beta_1$, \dots , $M^b_{\ell}=\beta_{\ell}$ is given by
\begin{eqnarray} P ( M^a_1=\alpha_1, \dots ,
M^a_{k}=\alpha_{k}, M^b_1=\beta_1, \dots  , M^b_{\ell}=\beta_{\ell}\Bigr)
& &  \nonumber \\  = \| \pi ^b_{\ell
,\beta_{\ell}}(t^b_{\ell})
\dots \pi ^b_{1,\beta_{1}} (t^b_{1}) \, \pi ^a_{k,\alpha_{k}}
(t^a_{k})
\dots \pi ^a_{1,\alpha_{1}} (t^a_{1}) \, \psi
\| ^2 .& & \label{measform}
\end{eqnarray}
Considering that under a multitime translation the
Heisenberg operators transform as
\begin{eqnarray*}  \pi^a_{j,\alpha}(t_a') & = &  U^a_{-t'_a}\, \pi
^a_{j,\alpha}\, U^a_{t'_a}  =  U^a_{\tau_a}\, \pi
^a_{j,\alpha}(t_a)\, U^a_{-\tau_a} \  = \  U_\tau \, \pi ^a_{j,\alpha}(t_a)\,
U_\tau^{-1} \\
 \pi^b_{j,\beta}(t_b') & =  & U^b_{-t'_b}\, \pi
^b_{j,\beta}\, U^b_{t'_b}  =  U^b_{\tau_b}\, \pi
^b_{j,\beta}(t_b)\, U^b_{-\tau_b} \ =  \  U_\tau \, \pi ^b_{j,\beta}(t_b)\,
U_\tau^{-1}  \end{eqnarray*}
and  the state transforms according to (\ref{Utaudef}), one sees that the
formula
(\ref{measform}) is in fact multitime translation invariant. In particular, the
predictions of the quantum measurement formalism are independent of
the frame of reference. Thus the quantum mechanical measurement formalism for a
system
which consists
of independent widely separated subsystems is multitime translation invariant.

Note also that the probability of obtaining the results $M^a_{i}=\alpha_{i}$,
$M^b_{j}=\beta_{j}$  {\it given}\/ the results $M^a_1=\alpha_1$, \dots ,
$M^a_{i-1}=\alpha_{i-1}$, $M^b_1=\beta_1$, \dots  , $M^b_{j-1}=\beta_{j-
1}$,
\[ \frac{ \| \pi ^b_{j,\beta_{j}} (t^b_{j})
\dots \pi ^b_{1,\beta_{1}} (t^b_{1})\,  \pi ^a_{i,\alpha_{i}}
(t^a_{i}) \dots \pi ^a_{1,\alpha_{1}} (t^a_{1})\,  \psi
\| ^2
}{ \| \pi ^b_{j-1,\beta_{j-1}} (t^b_{j-1})
\dots \pi ^b_{1,\beta_{1}} (t^b_{1})\,  \pi ^a_{i-1,\alpha_{i-1}}
(t^a_{i-1}) \dots \pi ^a_{1, \alpha_{1}} (t^a_{1}) \psi
\| ^2 } \]
can be conveniently expressed as
\[ \| \pi ^b_{j ,\beta_{j}} ( t^b_{j}) \,  \pi
^a_{i,\alpha_{i}} ( t^a_{i}) \psi_{\rm eff}  \| ^2  \]
with the ``collapsed wave function''
\[ \psi_{\rm eff} = \frac{  \pi ^b_{j-1, \beta_{j-1}} (t^b_{j-1})
\dots \pi ^b_{1, \beta_{1}} (t^b_{1}) \, \pi ^a_{i-1, \alpha_{i-1}} (t^a_{i-1})
\dots \pi  ^a_{1,\alpha_{1}}
(t^a_{1}) \, \psi }{\| \pi ^b_{j-1, \beta_{j-1}} (t^b_{j-1})
\dots \pi ^b_{1, \beta_{1}} (t^b_{1}) \,
\pi ^a_{i-1, \alpha_{i-1}} (t^a_{i-1}) \dots \pi ^a_{1,\alpha_{1}}
(t^a_{1}) \, \psi \| } . \]
(We find an analogous formula if we condition on a smaller initial segment.)

Within this framework an EPR experiment can be described---the subsystems,
while not explicitly interacting, are coupled by their common wave function
$\psi
(t_a,t_b)$---and one can explicitly see, for this two-time yet orthodox
model, that {\it the EPR-Bell nonlocality does not demand the existence of a
preferred frame of reference.}

Despite the presence of EPR-correlations, these do not permit the
transmission of ``signals'': {From} the results of measurements on system $a$
alone, one can draw no inference about the possible interventions on system
$b$---the kinds of experiments performed on system $b$. The crucial
assumption responsible for this property is the commutativity
(\ref{commute}).  In axiomatic quantum field theory the analogue of this
assumption, namely the commutativity of Heisenberg operators corresponding
to measurements in spacelike separated regions, is one of the fundamental
postulates, sometimes called ``local commutativity'' or ``microscopic
causality'' (see for example \cite{SW}). It conveys that experiments in
spacelike separated regions do not disturb each other, so that relativistic
causality is not violated.  However, it is important to recognize (as well
as all too rare) that EPR and Bell have shown that the quantum correlations
between observables for which ``local commutativity'' holds cannot in
general be explained by a local theory!

Bell has shown that the GRW model can also be formulated in a multitime
translation
invariant manner (\cite{Bell}, Chapter 22). Bell's result is sometimes regarded
as
indicating that the GRW theory is superior to Bohmian mechanics with respect to
the
problem of finding a Lorentz invariant extension. In the next section we show
that such
a conclusion is perhaps unfounded.

\subsection{A multitime translation invariant Bohmian theory}\label{3.2}

We formulate a multitime Bohmian theory that is invariant under multitime
translation. Consider a system consisting of $n$ widely separated
subsystems, as described in Section \ref{3.1}, with an $n$-time wave function
satisfying
(the analogue of) (\ref{sgl2}).
As usual, we shall for simplicity put
$n=2$. We shall denote again by $N_a$ and $N_b$ the particle numbers in the
subsystems and put $N=N_a+N_b$.
The beables of the multitime Bohmian theory are first of all the
usual beables of a Bohmian theory, namely the wave function, here the
two-time wave function, and the trajectories of the particle configuration
in the two subsytems, $Q_a(t)$ and $Q_b(t)$. The straightforward way to
formulate a multitime translation invariant Bohmian theory for the
evolution of these paths is to introduce as an additional beable a
synchronization: a path in two-time ${\rm I\!  R}^2$, i.e., an equivalence
class of maps $(T_a,T_b): {\rm I\! R}\to {\rm I\!  R}^2$, $s\mapsto
(T_a(s), T_b(s))$ differing only in their parametrization.  The
synchronization together with the subsystem trajectories
defines a {\it synchronized N-path}\/ in configuration-space-time
parametrized by $s$ \[ \Bigl( T_{a}(s),\, Q_{a}(s),\, T_b(s), \,
Q_b(s)\Bigr) =: Z(s) \] with $Q_{a}(s)\equiv Q_{a}(T_a(s)),\ Q_b(s)\equiv
Q_b(T_b(s))$.  We prescribe for the synchronized $N$-path the following
guiding equation
\begin{eqnarray} \frac { dT_a}{ds} = 1,&  & \frac {
dT_b}{ds} =1, \nonumber\\ \frac { dQ_a}{ds} = v_a^{\psi }(Z),&   &
\frac{
 dQ_b}{ds}=v_b^\psi (Z), \label{bohm2}
\end{eqnarray}
with $v_a^\psi $ and $v_b^\psi $ given as usual by
\begin{equation} \label{velfield2} v_a^\psi
 = {\mbox {Im}}\, \frac {\nabla _{q_a} \, \psi}{\psi}, \quad v_b^\psi =
 {\mbox {Im}}\, \frac {\nabla _{q_b} \, \psi}{\psi}.
\end{equation}
{\it The Bohmian theory given by the Eqs.\ (\ref{sgl2},
\ref{bohm2}, \ref{velfield2}) does not have a preferred ``frame of reference,''
and is obviously invariant under} $L_\tau $, i.e., if $(\psi , \, Z)$ is
a solution of (\ref{sgl2}, \ref{bohm2}), then so is $(\psi ',\, Z')=(\psi
\circ L_\tau ^{-1}, L_\tau \circ Z)$.
The parameter $s$ labels the synchronization with respect to which the
nonlocal interaction is mediated: The velocity of system $a$ at the
parameter value $s$ depends, through $\psi(t_a,q_a,t_b,q_b)$, upon the
configuration of the $a$-system at time $T_a(s)$---more precisely, upon
$Q_{a}(s)$ and $T_a(s)$---as well as on the configuration $Q_b(s)$ and the
time $T_b(s)$ of the $b$-system corresponding to parameter value $s$.  In
particular, the velocity ``field'' is a functional of the two-time wave
function
at the appropriate times.
Physical significance pertains
only to the synchronized $N$-path $Z({\rm I\! R})\subset {\rm I\! R}^{2+3N}$,
not
to the particular parametrization determined by (\ref{bohm2}). Thus,
just as with Eqn.\ (\ref{1partbohm}),  (\ref{bohm2}) is physically equivalent
to
all equations of the form $\displaystyle \frac {dZ}{ds} = A(Z)
(1,v_a^\psi (Z) ,1,v_b^\psi(Z))$ with arbitrary positive functions
$A$ on ${\rm I\! R}^{2+3N}$.

For the statistical analysis of this theory, it is natural to look for a
distinguished measure. As a consequence of (\ref{sgl2}), we have the two
identities which have the form of continuity equations
\begin{equation} \label{ce1}
\frac{\partial |\psi |^2 }{\partial t_a}+ {\rm div} _{q_a} j_a^\psi =0
\quad \mbox{or} \quad {\rm div} _{z_a} {J_a^\psi}=0 \end{equation} and
\begin{equation} \label{ce2}
\frac{\partial |\psi |^2}{\partial t_b}+ {\rm div} _{q_b}j_b^\psi =0 \quad
\mbox{or} \quad {\rm div} _{z_b} {J_b^\psi }=0 \end{equation} with
$J_k^\psi =(|\psi
|^2, j_k^\psi )$ and $j_k^\psi = |\psi |^2 \, v_k^\psi = {\rm Im}\,  (\psi ^*
\,
\nabla_k \, \psi )$, $k=a,b$. By analogy with the statistical analysis of
the usual Bohmian mechanics, it might at first glance seem appropriate to
seek a stationary measure for $Z$, i.e., for the dynamical system given by
Eqs.\ (\ref{bohm2}, \ref{velfield2}). The continuity equation for this
dynamical system, for a (continuously differentiable) density $f: {\rm I\! R}
\times
{\rm I\! R}^{2+3N} \to {\rm I\! R}$,
\begin{equation} \label{contz} \frac{\partial f}{\partial s} + {\rm
div} _{z_a} (fw_a^\psi ) + {\rm div} _{z_b} (fw_b^\psi ) =0 \end{equation}
with
$w_k^\psi := (1,v_k^\psi )$, is, by (\ref{ce1}) and (\ref{ce2}), solved
(trivially) by $f=|\psi |^2$, which is stationary with respect to the
synchronization parameter $s$.

Hence $|\psi |^2$ is certainly a distinguished measure on the space
${\rm I\! R}^{2+3N}$ of initial values for Eqs.\ (\ref{bohm2}). But it
is
not normalizable (by unitarity); moreover, for the dynamical system given
by Eqs.\ (\ref{bohm2}, \ref{velfield2}) there can be no density,
normalizable on ${\rm I\! R}^{2+3N}$, that is stationary with respect
to
the evolution parameter $s$---since a stationary measure for $Z$ yields a
stationary  marginal measure for $T_a$, and by the first of
Eqs.\ (\ref{bohm2}) all stationary measures for $T_a$ must be proportional
to the Lebesgue measure on ${\rm I\! R}$. In this regard it is also important
to
recognize that a general probability density $f(s)$ satisfying
(\ref{contz}), while defining a probability measure on $N$-paths, does not
itself directly correspond to any clear statistical property of this
ensemble of $N$-paths, such as the distributions of crossings discussed in
Section \ref{2}.

Recall now that $s$ labels the synchronization, and recall as well the
suggestion that ``quantum equilibrium'' should hold on
``simultaneity surfaces'' \cite{DGZ90}. Thus we proceed as
follows: We first fix initial values for the subsystem times
$T_a(0)=:s_0$ and $T_b(0):= s_0+h$. Then the evolution
equations for $T_a$ and $T_b$
may be solved to obtain
$T_a(s)=s_0+s$ and $T_b(s)=s_0+h+s$. The constant of motion $h=
T_b(0) -T_a(0)\, (=  T_b(s) - T_a(s)$ for all $s$) defines the
synchronization---the velocity of system  $a$ at a  time  $t_a$
depends upon the configuration $Q_b$ at time  $t_b=t_a+h$. (As it happens, to
this
fixed synchronization we may associate a (Lorentz) frame of reference in
which the interaction between the systems is ``instantaneous.''
However, this associated frame is merely a convenience, one that for more
than four  subsystems it would typically be impossible to retain.)

Now the subsystem times  in Eqn.\ (\ref{bohm2}) may be eliminated:
With
\[ \psi^{h}(s,q_a,q_b)  :=  \psi (T_a(s),q_a, T_b(s),q_b)  =  \psi
(s_0+s,q_a, s_0+h+s,q_b) \]
one obtains
\begin{eqnarray*} \frac { dQ_a}{ds} & = & v_a^{\psi ^h}(s,Q_a(s),Q_b(s)), \\
\frac{ dQ_b}{ds} & = & v_b^{\psi^h}(s,Q_a(s),Q_b(s)).\end{eqnarray*}
This is the usual Bohmian
mechanics relative to the synchronization given by $h$; we have the
continuity equation
\begin{equation} \label{konth} \frac{\partial {\rho^h}}{\partial s} + {\rm
div}_{q_a}
(\rho^h \, v_a^{\psi^h}) + {\rm div}_{q_b} (\rho^h \, v_b^{\psi^h})
=0,\end{equation}
and the
density $\rho^h = |\psi ^h|^2$ is ``equivariant,'' i.e., if $\rho ^h (s_0)
=|\psi^h (s_0)|^2$ for some $s=s_0$, then $\rho ^h (s)=|\psi ^h(s)|^2$ for
all $s$. For $\psi ^h(s)\in L^2({\rm I\! R} ^{3N})$, this density is
normalizable, and gives the distribution of crossings of any hypersurface
corresponding to the times $T_a(s)$ and $T_b(s)$ for the ensemble of
$N$-paths defined by $\psi ^h$.

\subsection{Hardy's experiment in multitime translation
invariant \protect\\ Bohmian theory}\label{3.3}

\begin{figure}[t]
\begin{center}
\leavevmode
\epsfxsize=10.3cm
\epsffile{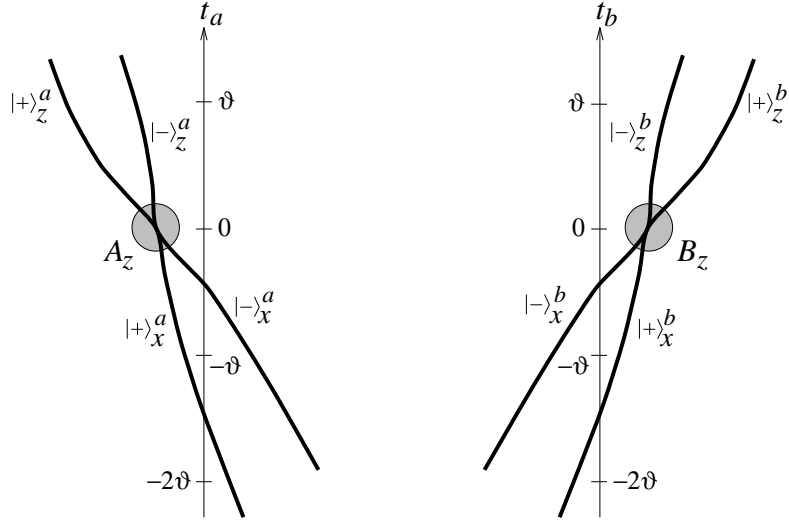}
\end{center}
\caption{Hardy's experiment in multitime formalism.}
\label{hardy2bild}
\end{figure}

We describe now the particle trajectories  in
Hardy's experiment for  the multitime translation invariant Bohmian theory
given by
Eqs.\ (\ref{sgl2}, \ref{bohm2}, \ref{velfield2}) (with $N_a=N_b=1$ and
the Stern-Gerlach magnets treated,  as usual, as external fields).
We prepare a system of two particles in the quantum state
$\psi_{\rm Hardy}$ (\ref{hardy1}). After the particles are widely
separated from each other, we perform  Hardy's experiment, cf.\
Figure  \ref{rhopsibild}, focusing on the part of the
experiment in which the particles run through the Stern-Gerlach magnets
$A_z$,
$B_z$, cf.\ Figure \ref{hardy2bild}.

\begin{figure}[p]
\begin{center}
\leavevmode
\epsfxsize=10.3cm
\epsffile{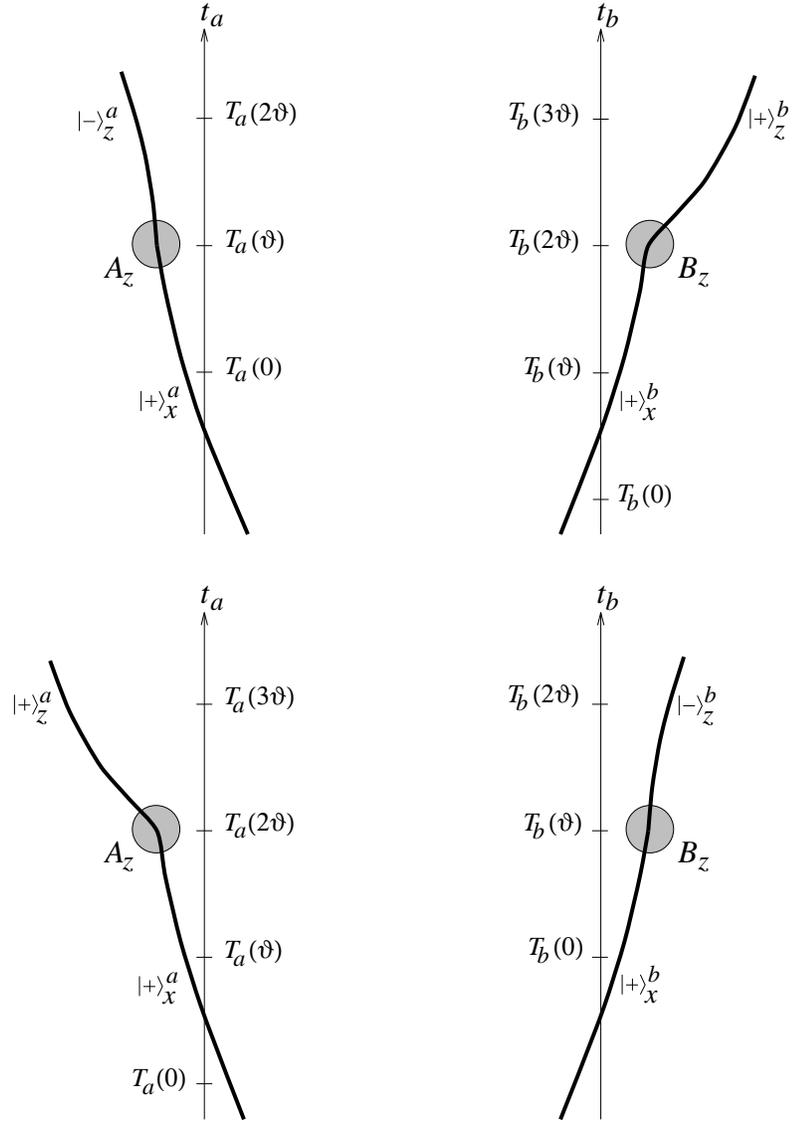}
\end{center}
\caption{Course of some synchronized two-paths according to the multitime
translation
invariant Bohmian theory in Hardy's experiment for initial values of subsystem
times
corresponding to the synchronization  {\it II}\/ (top) resp.\
{\it III}\/ (bottom).} \label{hardyIIbild}
\end{figure}

First we describe the development of the synchronized paths for initial values
of the
subsystem times $T_a(0)=-\vartheta$ and $T_b(0)= h^{\it II}=-2\vartheta$ for a
$\vartheta >0$,
not too large, referring to  scales for the subsystem times $t_a$ and $t_b$ as
defined in Figure \ref{hardy2bild}. This gives a synchronization corresponding
to the frame {\it II}\/ in
Figure \ref{rhopsibild}. Consider those two-paths for which at the value
$s=0$ of the synchronization parameter, particle $a$ is located in ${\rm
supp}\,
|+\rangle _x^a$ and particle $b$ is located in ${\rm supp}\,  |+\rangle _x^b$.
Demanding $\rho ^{h^{\it II}}(0)= |\psi ^{h^{\it II}}(0)|^2$, these are
1/12 of all two-paths. After particle $a$ has gone through the apparatus
$A_z$, it must be located in ${\rm supp}\,  |-\rangle _z^a$ since e.g.\ $\rho
^{h^{\it II}}(3\vartheta /2)= |\psi ^{h^{\it II}}(3\vartheta /2)|^2$, cf.\
Eqn.\
(\ref{psi2}). After particle $b$ has run through the apparatus $B_z$, it must
be
located in ${\rm supp}\,  |+\rangle _z^b$ since e.g.\
$\rho ^{h^{\it II}}(5\vartheta /2)= |\psi ^{h^{\it II}}(5\vartheta /2)|^2$,
cf.\
Eqn.\
(\ref{psi12}). This course of the particle paths is displayed in Figure
\ref{hardyIIbild}, top.

Now consider the same experiment with different initial values
for the subsystem times:  $T_a(0)=-2\vartheta$ and $T_b(0)=-\vartheta$, so that
$h= h^{\it III}=\vartheta$, a synchronization corresponding to the frame
{\it III}\/ in Figure \ref{rhopsibild}. Again consider those
two-paths  for which at the value  $s=0$ of the synchronization
parameter, particle $a$ is located in ${\rm supp}\,  |+\rangle _x^a$ and
particle  $b$ is located in ${\rm supp}\,  |+\rangle _x^b$. Demanding
$\rho  ^{h^{\it III}}(0)= |\psi  ^{h^{\it III}}(0)|^2$,  these are
1/12 of all two-paths. After particle $b$ has gone through  $B_z$, it must be
located  in ${\rm supp}\,  |-\rangle _z^b$ since e.g.\ $\rho
^{h^{\it III}}(3\vartheta /2)= |\psi  ^{h^{\it III}}(3\vartheta /2)|^2$, cf.\
Eqn.\ (\ref{psi3}). After particle $a$ has run through the
apparatus $A_z$, it must be located in
${\rm supp}\,  |+\rangle _z^a$ since e.g.\ $\rho  ^{h^{\it III}}(5\vartheta /2
)= |\psi  ^{h^{\it III}}(5\vartheta /2)|^2$, cf.\ Eqn.\ (\ref{psi12}).
This course of the particle paths is displayed in Figure
\ref{hardyIIbild}, bottom.

In neither case does the distribution of crossings by the two-paths of a
hypersurface corresponding to the other synchronization agree for all
parameter values $s$ with the corresponding $|\psi ^{h}(s)|^2$. In the
first case, $h=h^{\it II}=-\vartheta$, the two-paths run through ${\rm supp}\,
|+\rangle
_x^a \times {\rm supp}\,  |+\rangle _z^b$ when they cross a suitable
hypersurface
corresponding to frame {\it III}, even though the wave function $\psi ^h$ is
orthogonal
to $|+\rangle _x^a |+\rangle _z^b$, cf.\ Eqn.\ (\ref{psi3}).  Analogously,
in the second case, $h=h^{\it III}=\vartheta$, the two-paths run through ${\rm
supp}\,
|+\rangle _z^a \times {\rm supp}\,  |+\rangle _x^b$ when they cross a suitable
hypersurface corresponding to frame {\it II}, even though the wave function
$\psi ^h$ is
orthogonal to $|+\rangle _z^a |+\rangle _x^b$, cf.\ Eqn.\ (\ref{psi2}).

Finally we explain why, despite the fact that the two-paths (occasionally)
are in regions where the wave function vanishes, no violations of the quantum
mechanical
predictions would be experimentally observed. If an actual
experiment---involving, for example, the insertion of photographic plates
into the paths of the particles---were performed, the influence of this
apparatus on the future evolution of the complete system would have to be
taken into account. This can conveniently be accomplished, in a manner
analogous to what is done in ordinary Bohmian mechanics, by suitably collapsing
the wave
function $\psi ^h$ upon measurement. Suppose, for example, that we attempt to
detect
what quantum mechanically should be impossible, namely  the two-paths
running through ${\rm supp}\,  |+\rangle _x^a \times {\rm supp}\,  |+\rangle
_z^b$, which
we have just seen has positive probability for synchronization {\it II\/},
at least when no detection is attempted. We might do this by inserting a
detector in the path corresponding to ${\rm supp}\,  |+\rangle_x^a$, say at a
position corresponding to $s=0$, as well as in the path corresponding to
${\rm supp}\,  |+\rangle _z^b$. Then  with the synchronization {\it II}, the
wave
function $\psi
^{h^{\it II}}$ collapses at the synchronization parameter value $s=0$, when
particle $a$ is found in ${\rm supp}\,  |+\rangle _x^a$, to
\[ \psi _{a_x=+1}^{h^{\it II}} (s=0) = \frac 1{\sqrt 2}|+\rangle
_x^a\Bigl( |+\rangle _x^b-|-\rangle _x^b\Bigr) \
\stackrel{U_{5\vartheta /2}}\longrightarrow \ |+\rangle _x^a |-\rangle _z^b ,\]
and
the future evolution of particle $b$ changes drastically from what it would
have been like had there been no measurement or collapse: after having gone
through apparatus $B_z$ it no longer runs into the $|+\rangle _z^b$
channel, but rather into the $|-\rangle _z^b$ channel!  Analogous
things happen with the synchronization {\it III} and the opposite measurements.

\section{Reflections on Lorentz invariance and statistical
transparency}\label{4}

Concerning the model of Section \ref{3.2}, we have just alluded to the fact
that, just as for ordinary Bohmian mechanics, from the quantum equilibrium
hypothesis that
the actual distribution of crossings $\rho ^h=|\psi ^h|^2,$ one can derive
the quantum mechanical measurement formalism---which, as shown in Section
\ref{3.1}, is
multitime translation invariant and moreover does not even depend upon the
quantity $h$.
We thus have, with regard to our multitime Bohmian model, three levels of
description: the microscopic dynamical level, given by
(\ref{sgl2},\ref{bohm2}), which is multitime translation invariant; the
statistical
mechanical level, given by the quantum equilibrium hypothesis, which is, in
precisely the
same way, also multitime translation invariant---despite the results of Section
\ref{2};
and the observational level given by the quantum measurement formalism, which
is
also apparently multitime translation invariant.

There is, however, an important difference between the relativistic
characters of these levels: the latter level might be regarded as more
fully relativistic than the first two, which achieve their invariance
through the incorporation of the additional structure provided by the
synchronization. It might be argued that such a structure violates the
spirit of relativity \cite{Maud95,Maudlin}, and regardless of whether or
not we agree with this, it must be admitted that achieving relativistic
invariance in a realistic (i.e., precise) version of quantum theory without
the invocation of such structure seems much more difficult.  Hence Bell's
excitement about his version of the model of GRW (\cite{Bell}, Chapter
22). (It must also be admitted that a somewhat unpleasant implication of
the situation just described is that this synchronization structure---which
after all comprises a radical addition to physics---is, in the model under
consideration here, completely unobservable! See also \cite{Maud95}.)

Indeed, any theory can be made trivially Lorentz invariant (or invariant
under any other space-time symmetry) by the suitable incorporation of
additional structure, for example as given by the specification of a
Lorentz frame $\Lambda_0$ as part of the state
description.\footnote{Consider a theory specifying the set ${\cal L}$ (THE
LAW) of possible decorations $\xi$ of space-time and assume that the
Lorentz group acts naturally on any $\xi$ and thus on ${\cal L}$. This
theory, demanding that $\xi\in{\cal L}$, will be Lorentz invariant if
$\Lambda{\cal L}={\cal L}$ for any Lorentz transformation $\Lambda$.
Suppose this is not true. We may then enlarge the original theory by
replacing $\xi$ by $\hat\xi\equiv(\xi,\Lambda_0)$ and the law ${\cal L}$ by
$\hat{\cal L}$ defined by stipulating that ${(\xi,\Lambda_0)=\hat\xi\in
\hat{\cal L}} \Leftrightarrow {\xi\in \Lambda_0{\cal L}}$. (The original
theory thus corresponds to $\Lambda_0=I$.) Then $\hat{\cal L}$ is trivially
Lorentz invariant: For any Lorentz transformation $\Lambda$ we have that
$\Lambda\hat\xi=(\Lambda\xi,\Lambda\Lambda_0)\equiv(\xi',{\Lambda_0}')
\equiv\hat\xi'$, so that ${\hat\xi\in\hat{\cal
L}}\Rightarrow{\Lambda\xi\in\Lambda\Lambda_0{\cal L}}\Rightarrow
{\xi'\in
{\Lambda_0}'{\cal L}}\Rightarrow{\hat\xi'\in \hat{\cal L}}$.}  It seems
rather clear that this example, while Lorentz invariant, does not possess
what Bell has called ``serious Lorentz invariance,'' a notion, however,
that it is extremely difficult to make precise in an adequate way
\cite{Bell}.

The Bohmian model (\ref{bohm2}) immediately suggests a genuinely (though
perhaps not seriously) Lorentz invariant  Bohmian theory: For $N$ particles,
the beables are a multitime wave function and a synchronized $N$-path,
i.e., an equivalence class of maps $(X_1,\dots ,X_N): {\rm I\! R} \to
{\rm I\! R}^{4N}$, $s\mapsto (X_1(s),\dots ,X_N(s))$ differing only
in their parametrization. The synchronized $N$-path satisfies the guiding
equation
\begin{equation} \label{libohm2}
\frac{dX_k}{ds}  =  v_k(X_1(s), \dots , X_N(s)), \quad k=1,\dots ,N ,
\end{equation} where the
$v_k$ are  suitable 4-vector fields, on ${\rm I\! R}^{4N}$,
determined by the multitime wave function.
As with (\ref{1partbohm}) and (\ref{bohm2}), the fact that only the
synchronized $N$-path and not the parametrization determined by
a particular $v_k$ has beable status implies that all equations of the
form $\displaystyle \frac{dX_k}{ds}  =  a (X_1, \dots , X_N) v_k(X_1, \dots ,
X_N)$ with
an arbitrary positive function $a$ on ${\rm I\! R}^{4N}$ are
physically equivalent.

More concretely, one may consider a Lorentz invariant multitime Bohm-Dirac
theory: the wave function $\psi=\psi(x_1,\dots ,x_N)$ satisfies $N$ Dirac
equations
analogous to (\ref{sgl2}), and $v_k$ may for example be chosen to be
\begin{equation} \label{libohmdirac} v_k^\mu =  \bar \psi \gamma _k^\mu\psi
\end{equation}
with $\bar\psi = \psi^\dagger (\gamma^0\otimes \dots \otimes \gamma^0)
= \psi^\dagger \gamma^0_1 \dots \gamma^0_N$ and
$\gamma^\mu_k = \openone \otimes \dots \otimes \openone \otimes
\gamma^\mu\otimes
\openone \otimes \dots \otimes \openone$, the $\gamma^\mu$
at the $k$-th of the $N$ places.
We shall
discuss such a model in a subsequent work.  Just as with the model of
Section \ref{3}, models of the form (\ref{libohm2}), because of the nonlocal
interaction along the synchronization, have the possibility of properly
describing quantum nonlocality as exhibited, for example, by an EPR experiment.
This is in contrast with the local model of Squires \cite{sqmodel}, which
is based on what might be called a local light-cone synchronization. While
Squires formulates his model for the nonrelativistic Schr\"odinger
equation, he could as well have considered a multitime Dirac model with a
local light-cone synchronization to obtain a model that is completely Lorentz
invariant---and completely local.

Some readers may be wondering why we have analysed the nonrelativistic
multitime Bohmian theory in detail in Section \ref{3} instead of starting
right away with (\ref{libohm2}, \ref{libohmdirac}) or with the multitime
Bohm-Dirac theory
\begin{equation} \label{shbd} \frac{dT_k}{ds} = \psi^\dagger \psi , \quad
\frac{d{\bf Q}_k}{ds} = \psi^\dagger \bbox{\alpha} _k \psi \end{equation} with
$ X_k=
(T_k,{\bf Q}_k)$, $ k=1\dots N $, and $\alpha^i_k = \openone \otimes \dots
\otimes
\openone \otimes \alpha^i\otimes \openone \otimes \dots \otimes \openone =
\gamma_k^0
\gamma_k^i$.
This theory arises from Bohm's theory
(\ref{bohmdirac}) for $N$ Dirac particles $\displaystyle {\bf v} _k = \frac{
\psi^\dagger  \bbox{\alpha} _k \psi }{\psi ^\dagger \psi } = \frac{{\bf j}
_k}{ \rho}$ by introducing a dynamical synchronization, and it agrees  for
$N=1$
with (\ref{libohm2}, \ref{libohmdirac}). These models might suggest
that the reconciliation of statistical transparency and Lorentz invariance is
at hand.
However, for $N>1$ (\ref{shbd}) is not Lorentz invariant, because---unlike
(\ref{libohmdirac})---$( \psi^\dagger \psi , \psi^\dagger
\bbox{\alpha} _k \psi
)$ is not  a 4-vector.  On the other hand, (\ref{libohmdirac}) is not
statistically transparent because---unlike (\ref{shbd})---the
(reparametrization invariant)
configuration space velocity $v_k^i /v_k^0$ arising from (\ref{libohmdirac}) is
not of
the form ${\bf j}_k /\rho$ for $N>1$.  Thus, for the Lorentz invariant model
(\ref{libohm2}, \ref{libohmdirac}) equivariance does not hold in any
obvious way and hence, since there is in general no reason that quantum
equilibrium should hold with respect to any reasonable family of
hypersurfaces, the canonical statistical analysis cannot be performed and
the question of the extent of its agreement with standard quantum theory
becomes rather delicate.\footnote{This absence of statistical transparency  is
similarly also the case for the local model of Squires \cite{sqmodel}.}

There is another important difference
between (\ref{libohmdirac}) and (\ref{shbd}). To appreciate this consider the
system
\begin{equation} \label{multibd} \frac{dT_k}{ds} = 1, \quad \frac{d{\bf
Q}_k}{ds} = {\bf v} _k(X_1(s), \dots , X_N(s))  \end{equation}
with  $\displaystyle {\bf v} _k = \frac{
\psi^\dagger  \bbox{\alpha} _k \psi }{\psi ^\dagger \psi }$.
Here $(T_k(s))$ is
entirely determined by $(T_k(0))$ and the statistical analysis of this
theory may be developed as in Section \ref{3.2} for the multitime Bohmian
theory, merely replacing $|\psi |^2$ by $\psi ^\dagger \psi $.  With
(\ref{libohmdirac}), however, the equations for the evolution of the
synchronized particle times
\[ \frac{dT_k}{ds} = ( \bar \psi \gamma_k^0 \psi ) (X_1(s),\dots ,X_N(s)) \]
imply that in general $(T_k(s))$ depends upon the (initial) positions of
the particles as well as on $(T_k(0))$,  and it is difficult to see how one
could
begin any statistical analysis
even if the velocity field were otherwise somehow of a suitable form. Now
it might appear that we should have the same difficulty with (\ref{shbd});
however
the theory (\ref{shbd}) is equivalent to (\ref{multibd}) since the respective
vector
fields differ by a real-valued function on ${\rm I\! R}^{4N}$ and hence define
the same synchronized $N$-paths. Thus it turns out that
(\ref{shbd})
is  statistically transparent---or at least statistically
translucent.  We shall take up these
questions in a subsequent paper.

Observe that if the 4-vectors $v_k$ are 4-velocities ($v_{k\mu} v_k^\mu
=1$), the synchronization implied by
(\ref{libohm2}), which in this case is according to
proper time parametrization, reduces in the nonrelativistic limit to the first
set of
Eqs.\
(\ref{bohm2}).\footnote{Note that  $\bar \psi \gamma
_k^\mu\psi $ need not  in general be everywhere timelike and thus $v_k$
(\ref{libohmdirac}) cannot in general be normalized.
However, one can find a simple reparametrization $\displaystyle v_k=
\frac{\bar\psi \gamma_k\psi }{ \bar\psi \psi }$ such that the $v_k$
are approximate 4-velocities for ``large $c$'':
Writing $$\psi (x_1,\dots , x_N)=\sum _i \varphi_i (x_k)\chi_i (x_1,\dots ,
x_{k-1}, x_{k+1},\dots , x_N)$$ and noting that in the nonrelativistic limit
the last two components of $\varphi_i$ in the standard representation become
much smaller than the first two, one sees that in the nonrelativistic limit
$dT_k/ds=\bar \psi \gamma _k^0\psi / \bar\psi \psi \approx 1$ and the
space components of $v_k$ become small. Thus in the
nonrelativistic limit the theory (\ref{libohmdirac}) implies a
synchronization that can be
(re)expressed in the form (\ref{bohm2}) (first equations).
(Concerning the reparametrization by $\bar \psi \psi$ [and that by $\psi
^\dagger
\psi$ for the relation between (\ref{shbd}, \ref{multibd})], one
may convince oneself that not only the multiplication of the velocity field
by a positive function, but typically even by a fuction that has zeros or
changes sign will yield an equivalent theory.)}
Whatever reservations we may have concerning models such as
we've been discussing, a synchronization by proper time
seems to
us entirely compatible with serious Lorentz invariance, at least for a pair
of particles having a common origin in a single event.

The requirement that a Bohmian theory be Lorentz invariant without the
incorporation of such additional structure as a dynamical synchronization
places a very strong constraint on, say, the vector field defining the law
of motion (in a particular frame), or, what amounts to pretty much the same
thing, on the wave function of the system---and might be expressed via a
suitable fixed-point equation for this wave function. It seems extremely
likely that the set of wave functions satisfying such an equation is very
small, far smaller than the families of wave functions we normally consider
for the set of possible initial states of a quantum system.  However, if,
as is widely believed, we accept that from a cosmological perspective there
should be a unique wave function (for example, the Wheeler-de Witt wave
function or the Hartle-Hawking wave function) of the universe, this very
fact might well be a virtue rather than a vice!

\section*{Acknowledgements}
This work was supported in part  by
the DFG, by NSF Grant  No. DMS-9504556, and by the INFN.

\end{document}